# Metasurfaces for free-space coupling to multicore fibers

Jaewon Oh, Jun Yang, Louis Marra, Ahmed H. Dorrah, Alfonso Palmieri, Paulo Dainese, and Federico Capasso

*Abstract*— Space-division multiplexing (SDM) with multicore fibers (MCFs) is envisioned to overcome the capacity crunch in optical fiber communications. Within these systems, the coupling optics that connect single-mode fibers (SMFs) to MCFs are key components in achieving high data transfer rates. Designing a compact and scalable coupler with low loss and crosstalk is a continuing challenge. Here, we introduce a metasurface-based free-space coupler that can be designed for any input array of SMFs to a MCF with arbitrary core layout. An inverse design technique – adjoint method – optimizes the metasurface phase profiles to maximize the overlap of the output fields to the MCF modes at each core position. As proof-of-concepts, we fabricated two types of 4-mode couplers for MCFs with linear and square core arrays. The measured insertion losses were as low as 1.2 dB and the worst-case crosstalk was less than -40.1 dB in the O-band (1260-1360 nm). Owing to its foundry-compatible fabrication, this coupler design could facilitate the widespread deployment of SDM based on MCFs.

## I. INTRODUCTION

OPTICAL fibers are the foundation of communication systems today. Single-mode fibers (SMFs) have enabled high data transfer rates thanks to a variety of multiplexing techniques that encode information onto the dimensions of light, namely, its amplitude, phase, polarization, wavelength, and space. The first four dimensions of light are widely utilized in today's SMF-based systems and are expected to be exhausted in the near future. With current data demands increasing by an order of magnitude every four years [1], the impending "capacity crunch" has made space-division multiplexing (SDM), that is using the spatial degree of freedom of light, an active area of research over the past decade [2-7]. Multicore fiber (MCF) is a promising candidate for enabling SDM by parallelizing signal transmission through each of its cores [8-11]. However, backward compatibility with existing SMF-based multiplexing is necessary to expand the capacity of current systems. For this reason, the coupler that interfaces between SMFs and MCFs is a vital component for successful SDM-based communications.

Various techniques exist for designing MCF couplers, also known as fan-in fan-out (FIFO) devices, each with unique strengths and weaknesses. These can be categorized into three main classes: (i) free-space optics [12,13] (ii) 3D inscribed waveguides [14,15] and (iii) fiber-based devices [16-18]. The third type is realized by splicing a bundle of input fibers with a tapered cladding directly to the MCF. This method can provide very low loss since the coupler-MCF interface is essentially mode-matched and directly connected. However, precise geometry control for arbitrary core arrangement (linear, square, hexagonal lattices etc.) is particularly difficult. Furthermore, the assembly and tapering process presents challenges in scalability. 3D inscribed waveguides operate on a similar guided wave principle but are made by focusing a pulsed laser onto a dielectric medium, that locally modifies its refractive index. Using translation stages, waveguides can be written within the dielectric with high precision. The drawback is still scalability as writing waveguides is a serial process. In addition, this approach can only attain small index contrast (~0.36%) [14], limiting the achievable core pitch due to crosstalk. Couplers based on free-space optics use lenses to relay the input mode fields onto the MCF, enabling mode propagation with very low crosstalk (< -50 dB) and low loss (< 0.6 dB) [13]. The main disadvantage of this method is complexity. Each input SMF is connected to a collimator that must be precisely positioned and angled to a single lens that focuses each beam into their respective MCF core. Optional prisms placed between collimators and the final lens use beam displacement to reduce the coupler's footprint but introduces more sources of alignment error and this complication scales with the number of modes. A compact free-space design using a 4*f* system (two lenses) has been proposed [19] but it operates only by image magnification/demagnification and therefore does not allow

Manuscript received xx. This material is based upon research supported by Corning Inc. J. Oh, A. H. Dorrah, and F. Capasso acknowledge partial support from the Office of Naval Research under Award Number N00014-20-1-2450. This work was performed in part at the Harvard University Center for Nanoscale Systems (CNS); a member of the National Nanotechnology Coordinated Infrastructure Network (NNCI), which is supported by the National Science Foundation under NSF award no. ECCS-2025158. Part of the computations in this paper were run on the FASRC Cannon cluster supported by the FAS Division of Science Research Computing Group at Harvard University.

J. Oh, A. H. Dorrah, A. Palmieri and F. Capasso are with the Harvard John A. Paulson School of Engineering and Applied Sciences at Harvard University, Cambridge, MA 02138 USA. (e-mail: joh01@g.harvard.edu, capasso@seas.harvard.edu).

J. Yang, L. Marra and P. Dainese are with Corning Inc. in Painted Post, NY 14870, USA (e-mail: daineseP@corning.com).

coupling from an arbitrary arrangement of SMF arrays to arbitrary MCF core arrangements. For the same reason, the mode field diameter (MFD) at the output cannot be independently tuned thereby incurring loss due to mode mismatch. Therefore, the design of a compact and scalable coupler that can provide low loss and crosstalk is an on-going challenge.

In this paper, we report the first, free-space coupler for MCFs based on metasurfaces. Metasurfaces – planar optics composed of subwavelength-scale scatterers – have been used to engineer various aspects of light [20] including wavefront [21-24], dispersion [25], and polarization [26,27], making them an ideal platform for multiplexing optics [28]. The coupler construction is amorphous silicon (a-Si) nanopillars lithographically patterned on the front and back of a fused silica substrate. The operating principle of our device is like the bulk optics implementation, but it is achieved within the footprint of a *single* glass wafer. The metasurface patterns are generated using an inverse design technique called adjoint optimization to maximize the overlap of the device's output fields and MCF core modes. Based on this approach, we designed and fabricated two types of 4-mode couplers that take a linear 1-by-4 SMF array input and map them to (i) a linear 1-by-4 array output and (ii) a square 2-by-2 array output of MCF cores. The measured insertion losses were down to 1.2 dB and the crosstalk was up to -40.1 dB across the O-band (1260-1360 nm). We also provide a breakdown of different loss contributions for the fabricated 4-mode couplers to explain their origins and a roadmap for further improvement. Lastly, to demonstrate the versatility of our platform, we also simulated a 19-mode design for application in dense SDM and a broadband 4-mode design. We envision that our metasurface coupler could enable MCF-based SDM communications where high-throughput, scalability, and compactness are important requirements.

## II. DESIGN OF THE METASURFACE COUPLER

### A. Device concept

Fig. 1 depicts the schematic of the metasurface coupler. The incident light is launched from an array of SMFs to the metasurface located on the front surface of the device. The air gap distance between the input SMF-plane and this front surface is $f_i$. Each beam from the array is then deflected towards the metasurface located on the back surface, while propagating within a dielectric substrate of thickness $d$. This second metasurface cancels out the deflection and reshapes each beam to maximize the coupling to the MCF at the output plane. The air gap distance separating the back surface and the MCF is $f_o$.

It is important to note that these metasurfaces are not necessarily simple lensing elements that (de)focus the incident beams. The adjoint design approach *simultaneously* optimizes the phase profiles imparted by the front and back metasurfaces. As a result, all wavefront operations such as deflection, collimation, focusing, and mode-matching are co-optimized between these two planes. Compared to previous free-space coupling demonstrations [12,13,19], wavefront control with metasurfaces allows more design freedom such as arbitrary input/output beam arrangement, angle of incidence, propagation distance, *and* spatial mode profiles/MFDs of the MCF. Furthermore, this flexibility can be achieved at wafer-scale compactness due to the subwavelength thicknesses (in the $z$-direction) of metasurfaces.

### B. Design principle using adjoint optimization

Obtaining analytical phase profiles for the metasurfaces that minimize the losses of all modes is not straightforward under the constraints set by the device parameters. The key design challenge for the metasurface coupler is the following: given an incident field at some input plane and a target field at some output plane, determine a series of metasurfaces that maximizes the overlap between the launched incident field and the target field at the output plane. This is a well-known inverse design problem in photonics that can be addressed using a method called adjoint optimization [29-31]. (Note that this method is general and widely used outside of photonics, as well [32].) An objective function that describes the desired figure of merit (mode overlap, intensity at focus etc.) is formulated as a function of the fields. The device is expressed in the form of design variables (dielectric function, amplitude, phase etc.) that need to be optimized. Through a forward and backward propagation of the fields, the derivative of the objective function with respect to those design variables can be obtained and used in a gradient-ascent method to iteratively update the device until a satisfactory maximum for the objective function is reached. The phase profiles of our metasurfaces were optimized via this method. (If instead one desires that the objective function be minimized, gradient-descent is used.)

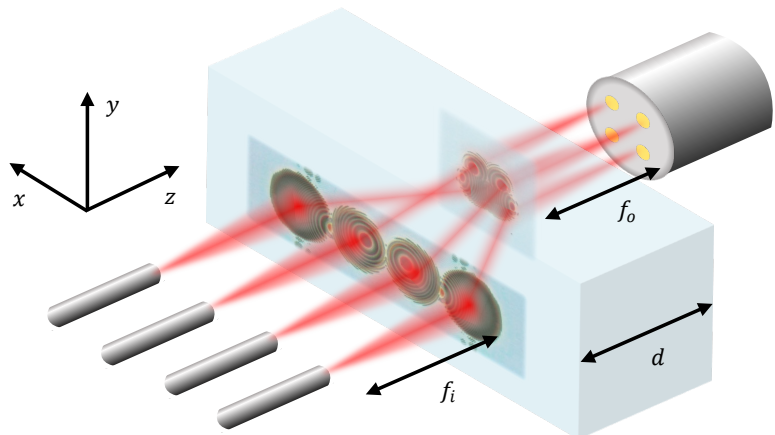

Fig. 1. Schematic of a metasurface coupler for SMFs to an MCF. Light is launched from an array of SMFs to the coupler. The metasurface on the front surface deflects the beams toward the metasurface on the back surface which then reshapes the deflected modes to maximize the overlap with the target MCF modes.

Fig. 2 shows the model of our adjoint optimization. Consider four planes defined along the $z$ direction: input, metasurface 1, metasurface 2, and output. The metasurfaces are treated as phase-only transmission masks $T^{(i)} = \exp(j\phi^{(i)})$ where $i = 1, 2$ defines the plane, $j^2 = -1$, and $\phi^{(i)}$ is the phase profile. Amplitude modulation is neglected to enforce unitary (i.e. no attenuation) wavefront shaping. For the $k^{th}$ mode, the launched field at the input plane is $u_k^{in}$ and the same mode as it propagates forward to the output plane is $u_k^{out}$. The target field that we want to couple to at the output plane is $v_k^{out}$. (All field quantities $u_k^{(i)}, v_k^{(i)}$ and phase profiles $\phi^{(i)}$ have dependence on $x$ and $y$ that is dropped for short-hand notation. Field propagation is modelled by Rayleigh-Sommerfeld diffraction theory.) To maximize the coupling of the launched mode to the desired target mode, the overlap integral of the two fields must be maximized at the output plane. This holds true for each $k^{th}$ mode and an objective function can thus be defined for all $K$ modes as

$$F = \frac{1}{K}\sum_k |\langle u_k^{out}|v_k^{out}\rangle|^2. \quad (1)$$

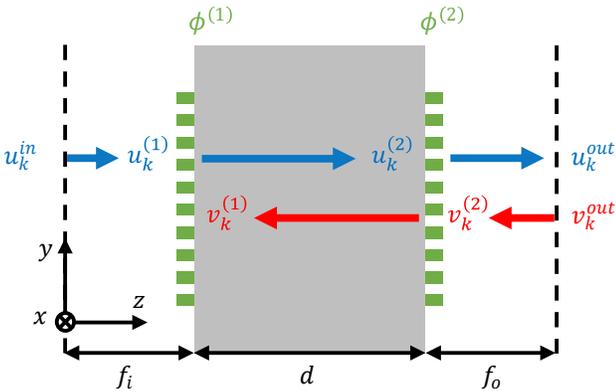

Fig. 2. Model of the metasurface MCF coupler used in the adjoint optimization. Two metasurfaces (green) are treated as phase-only transmission masks that are iteratively updated to maximize the overlap between the forward-propagating input fields (blue) and back-propagating target fields (red).

In the context of a SMF-to-MCF coupler, $u_k^{in}$ would be a SMF mode and $v_k^{out}$ would be the mode of a MCF core that we want to couple. In other words, the objective function (1) is just the mode-averaged insertion loss. Crosstalk, which is defined as the amount of power coupled into undesired cores or $|\langle u_k^{out}|v_m^{out}\rangle|^2$ for modes $m \neq k$, is another important figure of merit for mode couplers. However, it is not necessary to include it in the objective function (1) if low insertion loss can be achieved since most of the launched field will be localized at the position of the target core anyway. In mode converters where crosstalk is more prevalent, it has been shown that including a crosstalk term in the objective function can reduce it with negligible effect on insertion loss [33].

The gradient of (1) can be calculated with respect to the $i^{th}$ metasurface phase $\phi^{(i)}$ as

$$\frac{\partial F}{\partial \phi^{(i)}} = -\frac{2}{K}\sum_k \text{Im}\left[(v_k^{out\dagger}u_k^{out})^* v_k^{(i)*}\cdot u_k^{(i)}\right] \quad (2)$$

where Im is the imaginary part, † is the Hermitian adjoint, ∗ is the complex conjugate, and · indicates element-wise multiplication. $u_k^{(i)}$ is the forward-propagated field defined just *before* the $i^{th}$ plane and $v_k^{(i)}$ is the back-propagated (or adjoint) field defined just *after* the $i^{th}$ plane. A full derivation of the gradient can be found in [28]. Equation (2) states that for a given metasurface, the gradient can be obtained from just two simulations: the forward-propagated field and the adjoint field. The product of the terms inside the round brackets reduces to a constant. Thus, the physical interpretation of (2) is that the objective function $F$ changes proportionally to the mismatch between the forward and adjoint fields at the $i^{th}$ plane. This result is related to other phase-retrieval techniques such as the wavefront matching method [34] used for designing multi-plane light converters (MPLCs) [35,36]. The role of the metasurface is to then apply a spatial phase shift $\phi^{(i)}$ onto $u_k^{(i)}$ that corrects for the wavefront mismatch with respect to $v_k^{(i)}$. Equation (1) can be maximized by iterating $\phi^{(i)}$ with a gradient-ascent method of choice until convergence.

Although the adjoint optimization of the metasurface coupler allows full freedom over the device parameters, it does not guarantee the convergence of (1) to a large field overlap value. For example, if the device is constrained to a single transmission mask, beam displacement becomes physically impossible. Then, any input field $u_{in}^k$ and target field $v_{out}^k$ that is laterally displaced will experience high loss using a single-mask coupler. Moreover, the objective (1) is not for a single pair of input and target fields, but for $K$ pairs to be matched in limited space, making it further constrained. At the same time, adopting a large number of masks is not practical. The phase masks are modelled in the optimization as unitary but, in practice, the metasurfaces do not realize the masks with complete fidelity. The nanostructures composing the metasurface may reflect, absorb, or introduce phase errors which distort the desired wavefront transformation. As a result, it is beneficial to minimize the total number of masks needed in the coupler design. Consequently, our metasurface platform is comprised of no more than two surfaces; physically, this is sufficient to relay a mode from the input plane to any position on the output plane.

Once the optimized phase profiles are obtained, a look-up-table is used to map the nanopillar geometry, that best realizes the phase $\phi^{(i)}(x, y)$, to each location in the plane (see Appendix A). This defines the pattern of each metasurface. In principle, any technology platform (e.g., spatial light modulators, multi-level diffractive optics etc.) can be used to realize these phase masks as long as it is compatible with the parameters of the coupler model. In this regard, metasurfaces are advantageous for theoretical and practical (fabrication-related) reasons compared to other platforms. For the former, metasurfaces can support subwavelength pixel sizes which allow phase gradients large enough to direct a beam in any direction. The nanostructure within a pixel can also be designed to compensate for dispersion, enabling broadband performance [25,37]. From a fabrication standpoint, each metasurface only requires a single lithography step which is simpler than multi-level diffractive optics. For the couplers that were fabricated, we adopted nanopillars with circular cross-sections due to their low polarization dependence and high efficiency. The simulated amplitude and phase response of an a-Si nanopillar on a glass substrate at 1310 nm wavelength can be found in Appendix A. Full $2\pi$-phase coverage and high transmitted amplitudes can be achieved.

### C. Optimized phase profiles and simulated loss

Four different types of metasurface couplers were designed for the O-band centered at 1310 nm. Table I summarizes the model parameters for all couplers. The substrate is implemented as fused silica with an index of 1.4468. Only designs involving normally-incident SMF and MCF modes are explored as they are simpler to align in practice. The MCF cores here are all single-mode but couplers for multi-mode cores can be designed by defining the target field $v_k^{out}$ as a higher order mode.

The first two devices, Coupler A and Coupler B, demonstrate the flexibility over desired MCF core arrangements. While both take an input 1-by-4 linear array of SMF modes spaced 127 µm apart (like a fiber array unit), Coupler A maps them to a 1-by-4 linear array whereas Coupler B maps them to a 2-by-2 square array of MCF cores. The nearest-neighbor core pitches are 26 µm and 42 µm, respectively. The optimized phase profiles for the front and back metasurfaces are shown in Fig. 3a-d. The frontside patterns for both couplers resemble four lenslet-like hyperbolic phase profiles tilted to the direction of the respective MCF cores that they are designed to excite. Each lenslet roughly corresponds to where each input mode is incident. On the frontside of both couplers, we will refer to these as Modes 1 to 4, counting from left to right. The backside patterns are less intuitive but still appear lenslet-like. For Coupler A, the backside retains the same mode ordering as the front; however, for Coupler B, Modes 1 and 4 propagate through the bottom row and Modes 2 and 3 through the top row. Comparing the feature sizes on the front and back implies that the front narrows down the expanding beams from the SMFs. The tilted wavefronts are corrected back to normal incidence and the modes are re-focused to couple to the MCF. The corresponding simulated insertion loss $|\langle u_k^{out}|v_k^{out}\rangle|^2$ for each $k^{th}$ mode is shown in Fig. 4a-b. This is obtained by propagating the input field through the optimized phases to the output plane. Both devices are mirror-symmetric along the y-axis leading to the same loss curves for Modes 1 and 4 (outermost) and 2 and 3 (innermost). At 1310 nm extremely low loss (<0.2 dB) is possible, but this calculation assumes that the metasurface perfectly realizes the phase profiles and neglects Fresnel losses at each interface. Later, Coupler A and B were fabricated as proof-of-concepts for characterization.

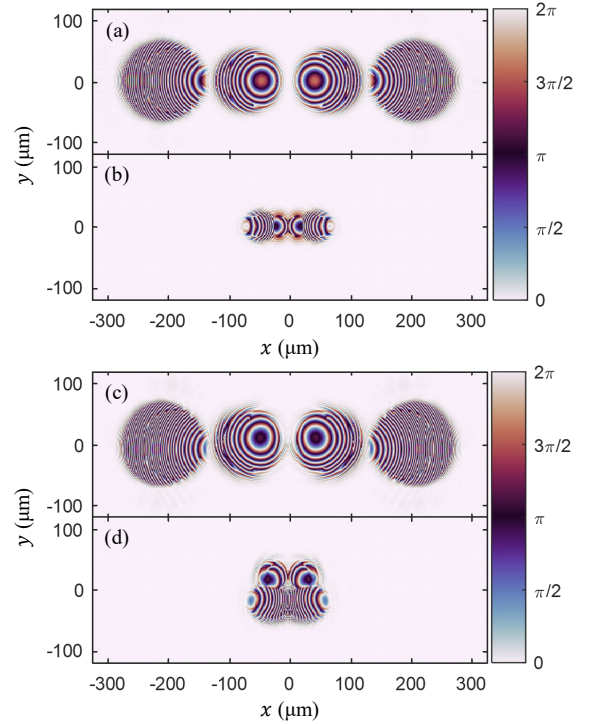

Fig. 3. Phase profiles of the adjoint-optimized metasurface couplers for two different MCFs. Phase profiles of the (a) frontside and (b) backside metasurfaces of Coupler A. Phase profiles of the (c) frontside and (d) backside metasurfaces of Coupler B. The front and back of the of the couplers face the SMFs and the MCF, respectively.

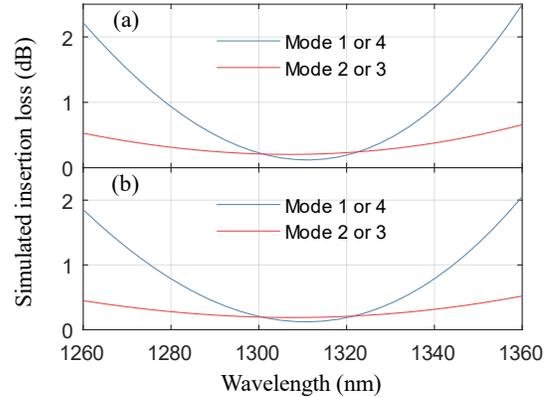

Fig. 4. Simulated insertion losses of adjoint-optimized metasurface couplers for two different MCFs. (a) Coupler A and (b) Coupler B.

TABLE I
DESIGN PARAMETERS OF METASURFACE COUPLERS

| Coupler design | SMF to device front spacing, $f_i$ (µm) | MCF to device back spacing, $f_o$ (µm) | Substrate thickness, $d$ (mm) | SMF input array | SMF array pitch (µm) | MCF core array | MCF core pitch (µm) |
|---|---|---|---|---|---|---|---|
| A | 345 | 70  | 0.8 | Linear 1-by-4 | 127 | Linear 1-by-4 | 26 |
| B | 345 | 105 | 0.8 | Linear 1-by-4 | 127 | Square 2-by-2 | 42 |
| C | 345 | 345 | 2.8 | Linear 1-by-4 | 127 | Linear 1-by-4 | 26 |
| D | 345 | 250 | 2.5 | Square 5-by-5 | 250 | Hexagonal 19  | 50 |

Coupler C is a broadband, 4-mode design with theoretical insertion losses below 0.7 dB over the entire O-band. It couples to the same type of MCF as Coupler A. The optimized phases and theoretical losses are shown in Appendix B. Compared to Coupler A, Coupler C differs in the second phase profile which resembles a single Fourier-transforming lens [12,13] with focal length $f_o$. The broader bandwidth can be attributed to the smaller bending angles which are less sensitive to changes in wavelength. Despite the improvement, this Fourier-transforming property requires the input SMF arrangement to match the output MCF core layout. Adding a third metasurface between the two can alleviate this limitation by acting as mode-dependent beam deflectors.

Coupler D tests the mode-scalability of the platform. It is a 19-mode coupler that maps a square input array of SMFs to a MCF with a hexagonal core array. The optimized phases and loss are also shown in Appendix B. It is essentially a larger version of Coupler A or B and unsurprisingly, the loss curves show resemblance.

### D. Fabrication

The wafer-scale compatibility of metasurfaces greatly simplifies their fabrication in contrast to other techniques. For example, to construct conventional lens-based couplers, every lens element must be precisely aligned relative to each other. As the mode count increases, the number of elements also increases and the alignment difficulty compounds. This problem does not exist in metasurfaces since all elements are lithographically written on a plane. Further, this allows high through-put scalability [38]. The only alignment involved, excluding the SMFs/MCF during measurement, is the positioning of the first metasurface relative to the second which is a standard process on lithographic writers.

The 4-mode couplers, A and B, were fabricated on the same fused silica substrate since they shared the same thickness parameter $d$ in the adjoint method. The a-Si films were grown on the top and back sides of the substrate. To ensure that the front and back metasurfaces would be aligned, gold (Au) crosshairs were first patterned on the front surface using a maskless aligner and lift-off process. Then, this was repeated on the back using the backside alignment feature of the maskless aligner such that the positions of the Au crosshairs matched those on the front. This alignment accuracy is better than 1 μm. The metasurface patterns were written and positioned relative to the crosshair positions, using e-beam lithography (EBL). Reactive ion etching (RIE) was used to etch the a-Si pillars. Full details of the fabrication steps are illustrated and described in Appendix C.

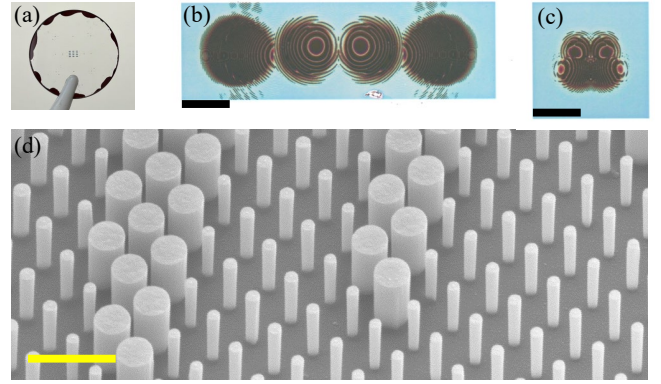

Fig. 5. Fabricated metasurface couplers. (a) A fused silica wafer of 1-inch diameter is shown with multiple devices patterned in the center. Both Couplers A and B are on the same substrate. Optical microscope images of the (b) frontside and (c) backside metasurfaces of Coupler B. The scalebar is 100 μm for both images. (d) Scanning electron microscope image of a region in (c) showing the a-Si nanopillars. The scalebar is 1 μm.

Fig. 5a shows the fabricated metasurface couplers on a 1-inch diameter substrate. To emphasize their compactness, there are twelve couplers (in a 4-by-3 grid) in the image; multiple copies were made for redundancy. An optical microscope image of Coupler B is depicted in Fig. 5b-c. The size of each coupler, excluding the input/output fiber spacing $f_i/f_o$, is no more than 650 μm by 240 μm by 800 μm; the last dimension is just the substrate thickness. A scanning electron microscope (SEM) image of the a-Si nanopillars that comprise the metasurface is shown in Fig. 5d.

## III. CHARACTERIZATION OF FABRICATED METASURFACE COUPLERS

### A. Output mode imaging

To first characterize Couplers A and B, the intensity profiles of the modes at the output plane were imaged. A schematic of the measurement setup is shown in Fig. 6a. A tunable laser source (Santec TSL-570) at 1310 nm wavelength was launched through a SMF (Corning SMF-28) mounted on a 5-axis stage (Suruga Seiki) with the three translation axes controlled by stepper motors and manual tip/tilt axes. The SMF tip is stripped bare and is positioned $f_i$ away from the front of the metasurface coupler at normal incidence. An objective lens (Mitutoyo NIR) collects the image at the output plane which is $f_o$ away from the back of the sample. Then, the image passes through magnification optics and is captured by an InGaAs camera (Allied Vision Goldeye). To image different modes, the SMF is simply scanned along the input plane according to the designed input spacing (127 μm) while keeping everything else fixed.

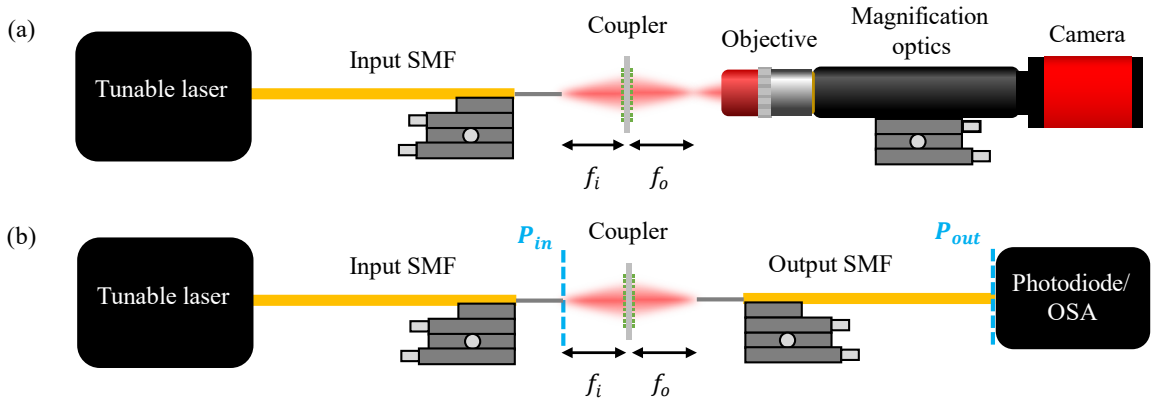

Fig. 6. Experimental setups for characterizing metasurface couplers for (a) mode imaging and (b) insertion loss and crosstalk. For (b) the SMF on the output side of the coupler was used to approximate the MCF since the modes are nearly identical. The stage moves the output SMF according to the positions of the MCF cores. A photodiode detector was used to measure the insertion loss. For crosstalk, an optical spectral analyzer was used instead for higher dynamic range.

The purpose of mode imaging was to verify the shape of the modes as well as the spacings between them. Instead of using the MFD as a merit, we calculated the mode amplitude overlap $\eta$ as a more informative way to measure the shape. For each mode, we define it as:

$$\eta = \left|\left\langle\sqrt{I_{meas}}\middle|v_{MCF}\right\rangle\right|^2, \quad (3)$$

where $I_{meas}$ is the spatial intensity profile of the output captured by the camera and $v_{MCF}$ is the analytical field of the MCF mode. Taking the square root of $I_{meas}$ approximates the field of the output image by dismissing its phase information. Each field quantity is normalized to have the same power. As a result, (3) ignores power losses and measures the mismatch of the two profiles in amplitude. This also quantifies information about stray light scattered away from the cores which cannot be obtained by simple MFD measurements. We emphasize that this calculation cannot be a direct estimate of loss since phase mismatch is excluded. Table II displays the amplitude overlap of each mode for each coupler at 1310 nm wavelength. These numbers assume that the total output field is captured by the objective and camera. The mode-averaged values for Coupler A and B are 94.1 % and 93.2%, respectively. To visualize how this relates to the mode shapes, Fig. 7a-b plots horizontal cuts of $I_{meas}$ against $|v_{MCF}|^2$ for the worst overlaps; they still match closely which means the metasurface is shaping the beam with high fidelity. Fig. 8a-b show composite images of the output plane at 1310 nm wavelength. One composite is made of four separate images taken for each mode at the same exposure time on the camera. The composites clearly show that the output mode spacings match the core pitches of 26 μm and 42 μm for Coupler A and B, respectively.

TABLE II
MEASURED MODE AMPLITUDE OVERLAPS (%) AT 1310 NM

| Coupler | Mode 1 | Mode 2 | Mode 3 | Mode 4 |
|---------|--------|--------|--------|--------|
| A | 94.6 | 94.4 | 93.0 | 94.5 |
| B | 94.0 | 92.6 | 92.5 | 93.9 |

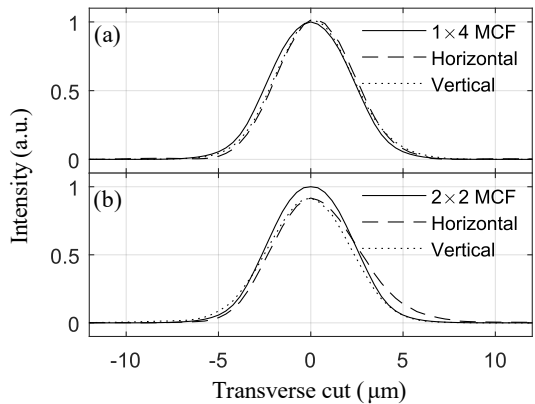

Fig. 7. Comparison of analytical MCF mode profiles (solid lines) and the measured intensity profiles (dashed and dotted lines) at the output plane of (a) Coupler A and (b) Coupler B. Mode 3, which has the lowest amplitude overlap for both devices, is shown.

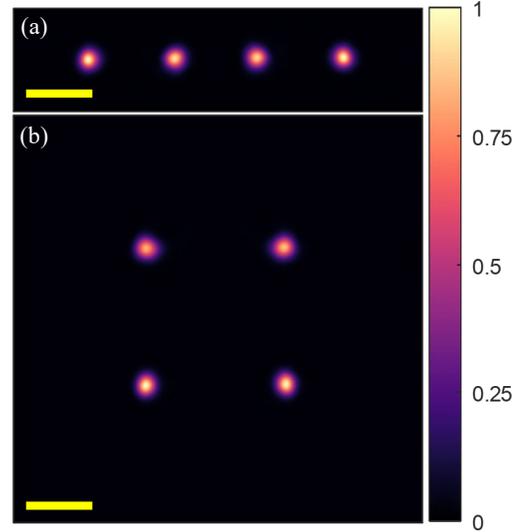

Fig. 8. Measured mode images at the output plane of (a) Coupler A and (b) Coupler B match the core positions of their respective MCF. This is a composite image made by scanning the input SMF and summing the output images for each mode. The positioning of the imaging optics and the camera exposure are fixed. Both scalebars are 20 μm.

*B. Insertion loss and crosstalk measurements*

Fig. 6b shows the setup used to measure the insertion loss and crosstalk. The setup preceding the metasurface sample is identical to the mode imaging setup. At the output side, another SMF (Corning SMF-28) with a bare tip is positioned normal to the backside of the device and $f_o$ away. This second fiber collects the mode at the output plane and is connected to a power detector at the other end. The output-side SMF approximates each core of the MCFs without the additional alignment difficulty introduced by the latter. This is justified since the SMF mode is nearly identical to the modes of the 1-by-4 and 2-by-2 MCFs (see Appendix D). The insertion loss is defined as the power measured coming out of the output fiber $P_{out}$ over the power measured at the input fiber $P_{in}$ using a photodiode detector. This assumes the propagation loss in the output fiber (~1 m) is negligible. We coiled the output SMF to verify that no power is carried in the cladding or high-order leaky modes. To measure each mode, the input SMF was scanned along the input plane as before, in addition to placing the output SMF in the output plane at the MCF core positions. Crosstalk data was taken using the same configuration as insertion loss except the photodiode was replaced with an optical spectral analyzer (OSA) (Yokogawa AQ6370D) for improved power dynamic range. Crosstalk is defined the same way as insertion loss except that, for each mode, the output fiber is moved to the core positions that we do not want to couple.

The measured insertion losses for Coupler A and B are shown in Fig. 9. The minimum loss over the O-band for both couplers is 1.2 dB. The losses for each mode at 1310 nm are shown in Table III. The corresponding mode-dependent loss (MDL) at the same wavelength are 0.1 dB and 0.2 dB for Coupler A and B, respectively. Because there are multiple, reflecting interfaces between the tip of the input fiber to the back facet of the output fiber, the insertion loss spectrum captures cavity-like oscillations. A Fourier transform of this spectrum (see Appendix E) reveals mainly three cavities formed between: (i) the back surface of the coupler and the facet of the output SMF (ii) the facets of the input and output SMFs and (iii) the facet of the input SMF and the front surfaces of the coupler. This effect can be eliminated with index-matching cladding encapsulating the metasurfaces but requires re-design of the nanopillars to account for the higher background index as well as additional fabrication optimizations. We also measured the polarization-dependent loss (PDL) for Coupler A at 1310 nm by introducing a paddle-type, fiber polarization controller between the laser source and the input SMF. We scanned the input polarization state and recorded the difference of the maximum and minimum insertion loss for each mode. The PDL was no greater than 0.1 dB for any mode.

TABLE III
MEASURED INSERTION LOSSES (dB) AT 1310 NM

| Coupler | Mode 1 | Mode 2 | Mode 3 | Mode 4 |
|---|---|---|---|---|
| A | 1.3 | 1.4 | 1.3 | 1.4 |
| B | 1.5 | 1.4 | 1.3 | 1.5 |

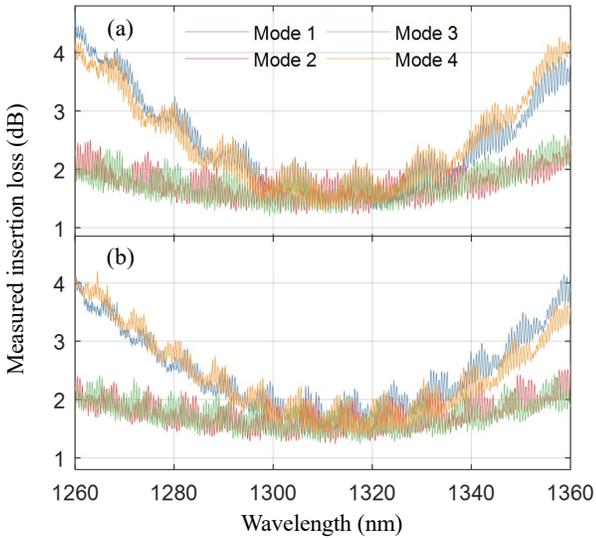

Fig. 9. Measured insertion losses of (a) Coupler A and (b) Coupler B for each mode across the O-band.

The crosstalk for all 4-mode combinations is plotted in Fig. 10. Unlike waveguide-based or fiber-based devices, the light inside the metasurface substrate is not guided. Thus, the crosstalk is due to some portion of light being diffracted outside of the intended target core at the output plane. For Coupler A, the worst-case crosstalk at 1310 nm and in the O-band is -46.3 dB and -40.1 dB, respectively. Unsurprisingly, these values are highest between nearest-neighbor cores (Modes 1 to 2, 2 to 3, and 3 to 4) as the undesirable diffracted power leaks not too far away from the target core position. The worst-case crosstalk for Coupler B at 1310 nm and across the O-band is -53.8 dB and -40.7 dB, respectively. It is expected that Coupler B has a lower crosstalk than A since the MCF core spacing is larger for the former. However, the crosstalk values away from the center of the band are similar for both devices.

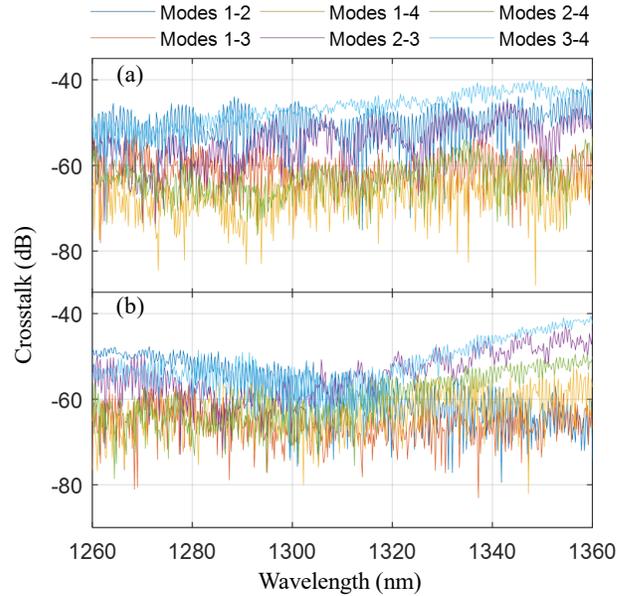

Fig. 10. Measured crosstalk of (a) Coupler A and (b) Coupler B for each mode pair across the O-band.

### C. Loss analysis

In this section, we break down the measured insertion losses for Coupler A into two major sources: transmittance and phase error. A comparison between the measured (Fig. 9a) and simulated (Fig. 4a) insertion losses suggests a relatively good agreement in the shape of the curves but with an offset of 1-2 dB. The following analysis reveals how much each source contributes to this extra loss and if they account for the total difference.

Recall that the simulated insertion loss assumed unitary transmission through each metasurface. To test its validity, a simple transmittance measurement can be conducted. The same setup as Fig. 6b is used except the output fiber is removed and the power detector is brought as close as possible to the output side of the metasurface coupler. This measures the total power that transmits through the device and enters the aperture of the photodetector. The light that is not captured is reflected by either metasurface, guided within the glass wafer, or transmitted beyond the angular range of the detector. Absorption is neglected since a-Si and fused silica is transparent in the O-band. Fig. 11 plots this transmittance loss for each mode. At 1310 nm, the values are 0.7 dB for Mode 1 and 0.6 dB for Modes 2 through 4.

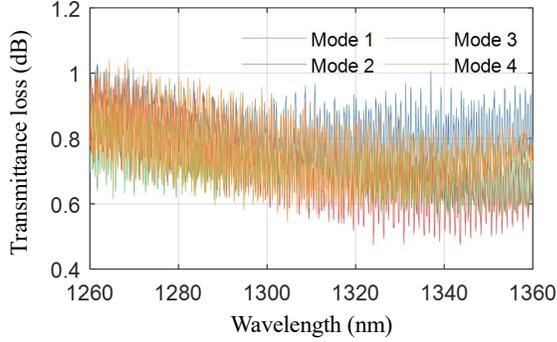

Fig. 11. Measured transmittance losses for each mode of Coupler A in the O-band.

The simulated loss also requires that the actual phase imparted by the metasurfaces matches the adjoint-optimized phase profiles (Fig. 3a-d). Any deviation leads to a wavefront error that contributes to the measured insertion loss. The actual phase of the device can be retrieved via digital off-axis holography [39]. A separate sample containing metasurfaces patterned only on one side of the substrate is placed on one arm of an interferometer. The pattern being probed is illuminated with a collimated beam and the transmitted light is collected by an objective lens. This light is interfered with a tilted reference beam and the interference pattern is recorded by a camera. The wavefront of the metasurface can be obtained digitally by isolating the off-axis Fourier component of the image, centering it, and taking the inverse Fourier transform [40]. Fig. 12 shows the measured phase profiles and phase error for Mode 2 of the frontside of Coupler A at 1310 nm wavelength. Modes 1 and 4 could not be accurately measured since their phases contain gradients above the NA-limit of our imaging optics; the backside pattern also contains these large gradients but only within a small region. The root mean square (RMS) wavefront errors for Modes 2 and 3 on the frontside and the full backside are $\lambda/14$, $\lambda/13$, and $\lambda/11$, respectively. Note that in this calculation, circular apertures were cut-out of the phase profiles to reduce the effect of the uniform, low-phase region which would otherwise deflate the error.

A loss can be correlated with this phase error by modelling the propagation of the input SMF mode through the measured phases. Each metasurface is treated as unitary. The overlap of this field at the output plane with the analytical MCF core mode yields this extra loss plus the simulated loss. For Mode 2, the result is 1.0 dB and the portion due to phase error is 0.8 dB. Ideally, the sum of the simulated loss, phase error contribution, and transmittance loss (1.7 dB) should equate to the measured insertion loss at 1310 nm (1.4 dB). The difference of 0.3 dB could arise from the missing high-NA component of the measured backside phase, the cavity effect between the device substrate and the output SMF fiber, and redundancies in transmittance and phase error loss since guided/forward-scattered light beyond the detector is counted in both. Despite these simplifications, the discrepancy is small and most of the insertion loss seems to come from both sources with comparable contribution. For the rest of the O-band, we expect the phase error to increase since the a-Si nanopillars were designed at 1310 nm wavelength.

Reducing both transmittance and phase error loss involves further design and fabrication considerations. First, part of the transmittance loss is caused by reflection at either metasurface. This is expected since the library of a-Si pillars (Appendix A) shows transmission amplitudes varying from 0.9 to 1 depending on the pillar diameter; the rest of the light is reflected. These nanopillars behave as truncated waveguides with each facet having some reflectivity due to the index mismatch of the propagating waveguide mode and air/substrate. Reflections can be lowered by designing a library with facets flanked by materials that are better index-matched to the propagating mode of the pillars. This can be partly realized by a leave-on, oxide mask sitting on top of the a-Si. Second, the loss due to phase error is a result of fabrication defects and limitations of library-based metasurfaces. Geometric distortions from the intended pillar diameters introduced during fabrication contribute to this error and can be refined by fine tuning e-beam patterning. On the design side, the library only considers interactions between identical pillars, while the actual device includes distinct neighboring pillars. Such interactions impact both the amplitude and phase response at each meta-atom and must be taken into account using full device optimization (although this is currently a computational challenge given the large area of our device) [41,42].

## IV. CONCLUSION

We report the first demonstration of a metasurface-based free-space coupler between SMFs and MCFs. These devices are fabricated with foundry-compatible processes that allow high-throughput scalability beyond current MCF coupler platforms. Metasurface couplers operate similarly to couplers based on free-space optics. However, the former offers more design degrees-of-freedom such as coupling from arbitrary fiber arrays to arbitrary MCF core arrangements. This is made possible given the wavefront manipulation capabilities of metasurface at sub-wavelength spatial resolution. Two different versions of metasurface couplers were fabricated and had measured insertion losses down to 1.2 dB in the O-band. The worst-case crosstalk was measured at -40.1 dB. In addition, two more designs were proposed to improve the bandwidth (< 0.7 dB over the O-band) and to test mode-scalability (19 modes), respectively. The loss in the fabricated device seems to stem from both transmittance and phase error loss. The former can be addressed in part by intermediate index-matching materials to reduce Fresnel reflections at the a-Si nanopillars and the latter with the optimization of a-Si nanostructures at the scale of the entire device. All of the demonstrated performance is achieved within the footprint of a single glass wafer.

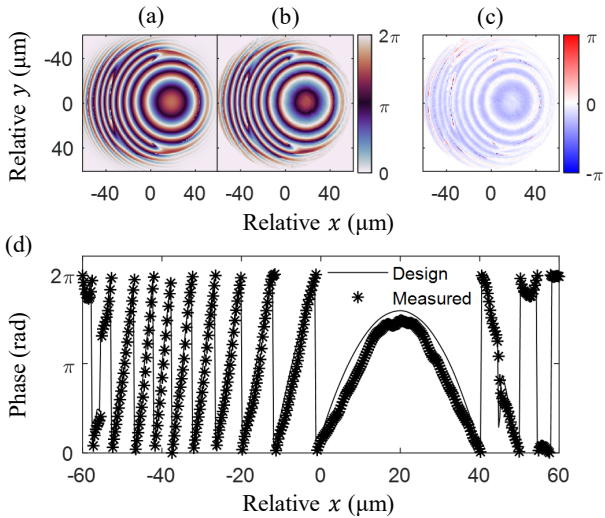

Fig. 12. Phase error analysis of Coupler A for Mode 2 of the frontside metasurface. (a) The design phase profile from adjoint optimization and (b) the measured phase profile obtained from digital off-axis holography. (c) The error between the measured and the design. (d) Horizontal cuts of (a) and (b) at $y = 0$. The relative positions of the phases are shown.

## APPENDIX

### A. Metasurface nanopillar library

Fig. 13 depicts the far-field amplitude and phase response of an a-Si nanopillar on top of a fused silica substrate for various diameters. The height of the pillars is 790 nm, the pixel size is 550 nm, and the incident wavelength is 1310 nm. Finite-difference time-domain (FDTD) (Ansys Lumerical) method is used to model the nanopillars and periodic boundary conditions are assumed. Once the response is known, the nanopillar geometry that best realizes the device phase profiles $\phi^{(i)}(x,y)$ (refer to Fig. 3a-d) is placed at that pixel for each position in the design.

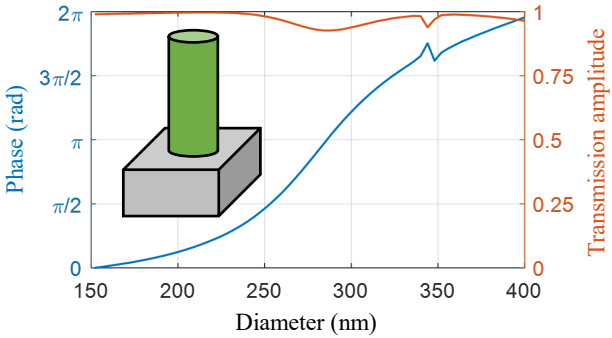

Fig. 13. Far-field transmission amplitude and phase response of a-Si nanopillars at 1310 nm wavelength. The height of the pillars is 790 nm and the pitch is 550 nm.

### B. Additional metasurface MCF coupler designs

Fig. 14 and Fig. 15 show the optimized phase profiles and simulated insertion losses for Coupler C and D, respectively. For Coupler D, the input array is a 5-by-5 square rotated 45 degrees; only the innermost 19 modes are launched. Since the device has symmetry, only the unique modes are shown in the insertion loss.

### C. Fabrication of metasurface MCF couplers

Out of the four coupler designs, Coupler A and B were the only two fabricated since the device parameters were closest to existing optimized recipes at our cleanroom facilities. Coupler C and D both have thicker substrates which impact thermal equilibration during resist baking and cooling during etching. This can be addressed by adjusting baking times and substrate cooling temperatures, respectively.

The fabrication steps are illustrated in Fig. 16. First, 790 nm-thick a-Si films were deposited on the front and back surfaces of the fused silica wafer (JGS2 grade) using plasma-enhanced chemical vapor deposition (PECVD). Four Au crosshairs, centered at the middle of the wafer, were patterned on the frontside using lift-off. Using a maskless aligner equipped with a backside alignment camera, four more Au crosshairs were patterned on the backside with lift-off again. The frontside metasurface pattern was processed first. A negative e-beam resist (ma-N 2400) was spin-coated and baked. A charge dissipating solution (Espacer) was also spin-coated on top of the resist. The phase profile pattern ($\phi^{(1)}$) was exposed using a 50 kV e-beam lithography writer (Elionix HS50) and was developed in AZ 726 developer. Espacer is water-soluble so it is removed during development. The pattern on the resist layer was transferred to the a-Si film via RIE (SPTS Rapier) using a simultaneous mixture of $SF_6$ and $C_4F_8$ gasses. (Although not depicted in Fig. 16c and f, the e-beam resist also masks the Au markers.) The remaining resist is removed using an oxygen plasma asher. With the frontside processing finished, we protected it by spin-coating a few layers of a temporary, polymethyl methacrylate (PMMA) film. The backside pattern ($\phi^{(2)}$) is then fabricated using the same steps as the front. Finally, the temporary film is removed using the same oxygen plasma asher.

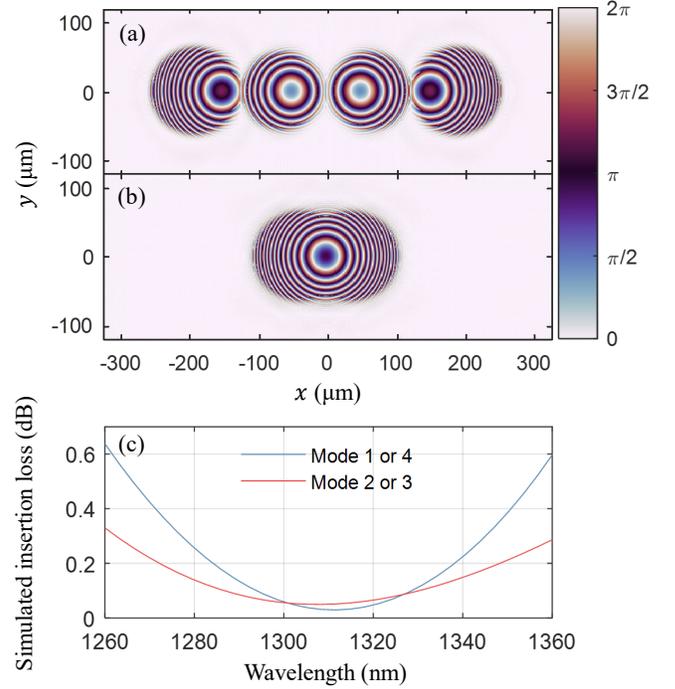

Fig. 14. Optimized phase profiles of the front (a) and back (b) metasurfaces of Coupler C. (c) The simulated insertion loss.

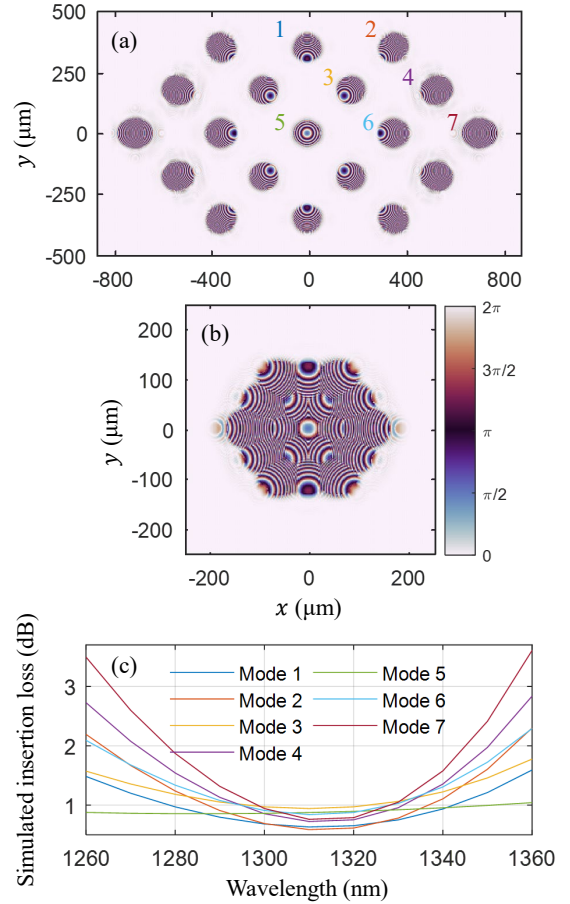

Fig. 15. Optimized phase profiles of the front (a) and back (b) metasurfaces of Coupler D. (c) The simulated insertion loss is shown. The device has mirror symmetry so only the unique modes are shown in the loss. The color coordination of the modes in (a) matches (c).

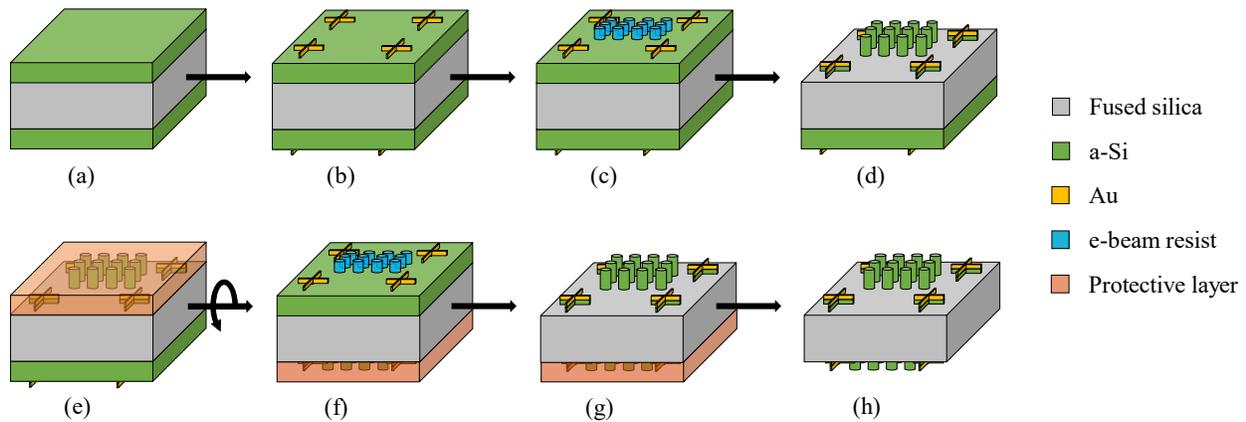

Fig. 16. Fabrication steps for metasurface MCF couplers. (a) PECVD of a-Si films. (b) Au marker patterning on both sides using lift-off. On the front, (c) pattern the e-beam resist via EBL and (d) transfer the pattern onto the a-Si layer via RIE. (e) Coat the finished frontside with a temporary, protective film. On the back, repeat the same steps used to pattern the a-Si using (f) EBL and (g) RIE. (h) Remove the protective layer on the front.

### D. Approximating MCF cores with a SMF

The measured insertion loss and crosstalk reported in Fig. 9 and 10 used a SMF to launch the input beam and another SMF to collect the mode at the output plane of the devices. Here, the SMF on the output-side is behaving like the cores of the MCFs by moving it to the core positions. Assuming negligible error from stage accuracy, this approximation is justified since the SMF mode is essentially the same as the core mode of both the 1-by-4 MCF and the 2-by-2 MCF. Fig. 17 compares their analytical field amplitude. The loss from mode mismatch between the core of either type of MCF and SMF is calculated to be no more than 0.02 dB at 1310 nm.

### E. Fourier transform analysis of measured insertion loss and transmittance loss

Fig. 18a plots the Fourier transform of the measured insertion loss of Mode 1 for Coupler A and B (from Fig. 9) versus the cavity length in air. A Fabry-Perot model is assumed and the optical path length of the glass substrate in air $d_{air}$ is 1157 μm. For both devices, a peak is seen around 345 μm which agrees with the input SMF-to-front metasurface gap $f_i$. Sharp peaks are located at 70 μm and 105 μm, corresponding to the back metasurface-to-output SMF gap $f_o$ for Coupler A and B, respectively. The broad peaks at long lengths are explained by reflections between fiber-to-fiber ($f_i + f_o + d_{air}$) and partly by input SMF-to-back metasurface ($f_i + d_{air}$). Another weak peak exists around 750 μm which could be a second order effect of the cavity with length $f_i$.

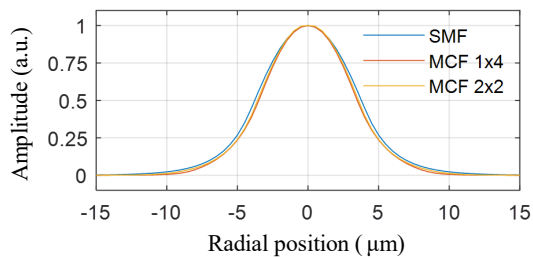

Fig. 17. Analytical mode profiles of the SMF, 1-by-4 MCF, and 2-by-2 MCF at 1310 nm.

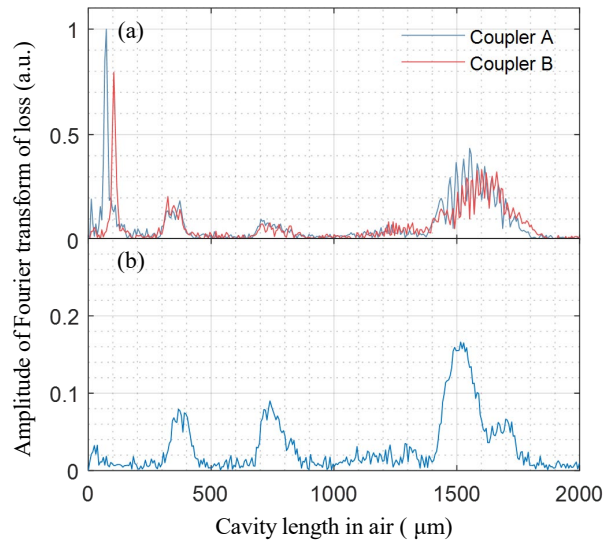

Fig. 18. (a) Fourier transform of the measured insertion loss of Mode 1 for Coupler A and B. (b) Fourier transform of the transmittance loss of Mode 1 for Coupler A.

Fig. 18b plots the Fourier transform of the transmittance loss of Mode 1 for Coupler A (from Fig. 11). Since the output SMF is replaced by a photodiode in the latter, the peak corresponding to $f_o$ is gone. The broad peak at long lengths is also centered slightly lower than in the insertion loss as only the input SMF-to-back metasurface cavity ($f_i + d_{air}$) exists.

ACKNOWLEDGMENT

J. Oh thanks J. S. Park for sharing the oxygen plasma ashing technique and A. Rezikyan for scanning transmission electron microscope (STEM) images and energy dispersive spectroscopy (EDS) analysis of fabricated samples. The authors would also like to thank Z. Yu (Flexcompute) for the invaluable opportunity to use the high-speed FDTD solver Tidy3D [43] which helped confirm fabrication issues with the first iteration of the device.